\documentstyle[prb,aps,multicol,epsfig]{revtex}

\begin{document} \bibliographystyle{unsrt}
\input{epsf}
\title{Magnetic Relaxation Phenomena in a CuMn Spin Glass}
\author
{
C. Djurberg, K. Jonason and P. Nordblad
}
\address
{
Dept. of Materials Science, Uppsala University, Box 534, S-751 21 Uppsala,
Sweden
}

\maketitle

\begin{abstract}
Experiments on the temperature and time dependence of the response
function and the field cooled magnetisation of a Cu(Mn) spin glass at
temperatures below the zero field spin glass temperature are used to
explore the non-equilibrium nature of the underlying spin
configuration.  The results imply that a certain spin configuration is
imprinted on the system as the temperature is decreased at a constant
cooling rate.  The cooling rate governs the magnitude of the FC
magnetisation ($M_{FC}$($H$,$T$)).  Any intermittent halt at a
constant temperature, $T_{i}$, imprints an extended spin
configuration, a process that is reflected e.g. in a downward relaxation of
$M_{FC}$.  On continued cooling at the same rate, the magnitude of
$M_{FC}$($T$) remains at a lower level than that of a continuous
cooling curve.  These results are put into the context of the
corresponding behaviour of the response function as observed in
measurements of the relaxation of the zero field cooled magnetisation.
\end{abstract}

\vskip 0.2cm

\centerline{PACS numbers: 75.40.Gb 75.50.Lk}
\vskip 0.2cm

\bigskip
\begin{multicols}{2}
\narrowtext

\section{introduction}

The non-equilibrium character of the slow relaxation of the
magnetisation of 3d spin glasses below the zero field phase transition
temperature has been extensively studied by dc-magnetisation
\cite{one,two} and ac-susceptibility \cite{three,four} experiments as
well as MC-simulations \cite{five}.  Different models to describe the
spin glass phase and the observed ageing phenomenon have been
suggested \cite{six,seven} and discussed in connection with the
empirical data.  In this paper we report results from dc-magnetic
relaxation experiments of the field cooled (FC) and the zero field
cooled (ZFC) magnetisation.  The field cooled magnetisation is found
to approach a reversible magnetisation level ($M_{FCrev}(T)$) if the
sample is continuously cooled and re-heated at one and the same rate
while remaining at temperatures below $T_{g}$.  On an intermittent
stop at constant temperature, the field cooled magnetisation relaxes
downward and on continued cooling the magnetisation remains at a lower
level than in a continuous cooling process.  On re-heating the sample
the lower magnetisation level is maintained only up to the temperature
of the intermittent halt, whereas on further heating above the halt
temperature the reference level of a continuous process is
progessively regained.

We also show that the wait time dependence of the response function at
constant temperature is governed by the cooling/heating rate and is
independent of any long time equilibration at sufficiently lower or
higher temperatures.  However, for temperatures that are only slightly
different, the response function is affected by a wait time at the
nearby temperature; a region of overlap is observed.

This behaviour suggests that there is a certain favorable
magnetisation associated with the spin configuration that the FC spin
glass attains in a cooling process at a specific rate.  This spin
configuration and its magnetisation remains essentially unperturbed if
the system is re-heated at the same rate.  If the sample is kept at a
constant temperature, the spin configuration is free to re-configure
unrestrictedly on large length scales and the FC-magnetisation
decreases.  This extended spin structure becomes imprinted and remains
'frozen' in when the sample is further cooled, and is only washed out
when the temperature reaches well above the temperature for the
intermittent halt.

\begin{figure}

\centerline{\hbox{\epsfig{figure=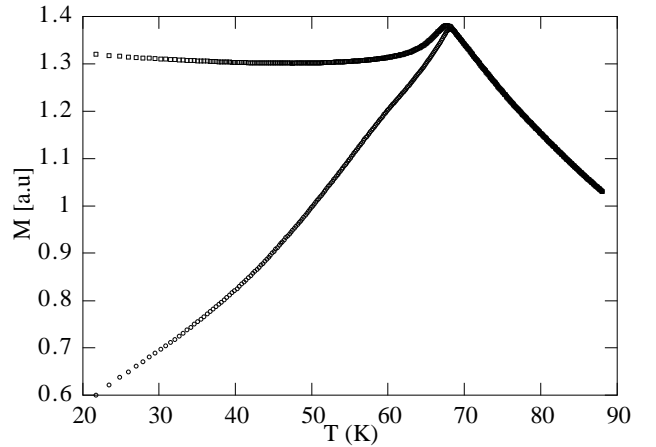,width=8.5cm}}}
\vskip 0.5cm
\caption{
\hbox { FC- and ZFC-magnetisation vs.
temperature} of Cu(Mn13.5at\%), $H$ =1 Oe.
}
\label{fig1}
\end{figure}

\section{experimental}

The sample is a bulk piece of a Cu(Mn13.5at$\%$).
The experiments were performed in a non-commercial SQUID magnetometer
specially designed for low field dynamic magnetic susceptibility
studies \cite{eight}.  Two basic experimental procedures are employed
in the study.

(i) Measurements of the zero field cooled magnetisation (the response
function).  The sample is cooled in zero field and subjected to
certain thermal sequences after which a weak magnetic field is applied
and the relaxation of the magnetisation is recorded at constant
temperature, i.e.  the
temperature and history dependent response function is measured.

(ii) Measurements of the field cooled magnetisation.  The magnetic
field is applied at a temperature well above $T_{g}$($H$=0) and the
sample is cooled (and also re-heated) in constant field.  The
temperature dependence, $M_{FC}$($T$,$t_{c}$) or the relaxation
$M_{FC}$($T_{m}$,$t$) of the magnetisation at constant temperature is
measured.  Here $t_{c}$ is a characteristic time determined by the
specific cooling/heating rate used in the experiment and $T_{m}$ is
the temperature for a relaxation measurement.

To acquaint with the sample the field cooled and zero field cooled
magnetisation is plotted vs.  temperature in Fig.  1.  The curves are
measured in a field of 1 Oe and show a cusp in the ZFC susceptibility
and an onset of irreversibility at about 68 K, which also closely
reflects the spin glass temperature, $T_{g}$, of the sample.

\section{response function}

\begin{figure}

\centerline{\hbox{\epsfig{figure=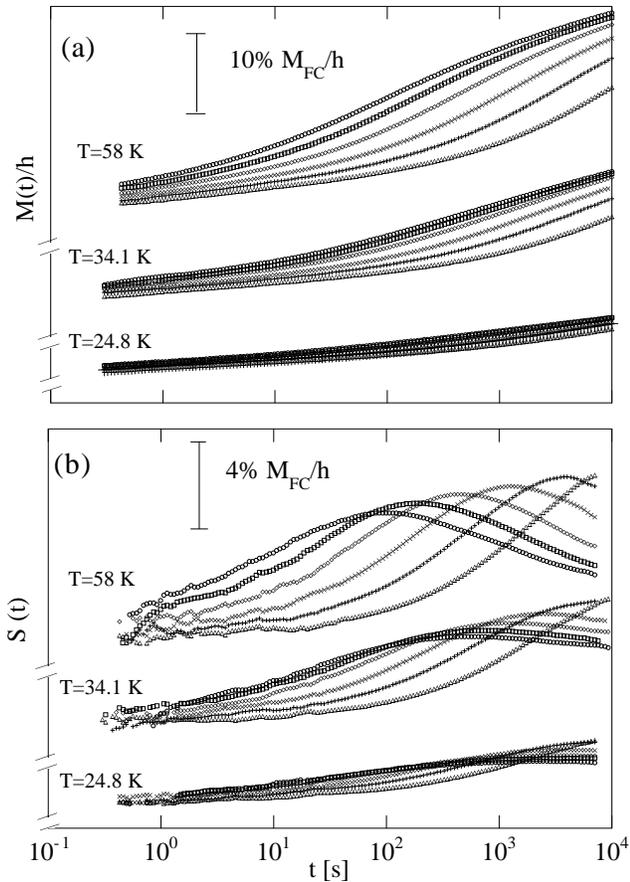,width=8.5cm}}}
\vskip 0.5cm
\caption{
\hbox { a) $M_{ZFC}$($T_{m}$,$t$) measured at three different}
temperatures and wait
times, $t_{w}$=30, 100, 300, 1000, 3000, 10000 s, b)
 the corresponding relaxation
rate, $S=1/H$ $\partial M$$/$$\partial$log$t$, $H$ =1 Oe.}
\label{fig2}
\end{figure}

Fig.  2 shows the zero field cooled magnetisation and the
corresponding relaxation rate, $S=1/H$ $\partial M$$/$$\partial$log$t$,
measured at three significantly different temperatures as a function
of wait time, $t_{w}$, at constant temperature before the magnetic
field is applied.  The wait time dependence of the magnetic relaxation
is somewhat different at the three temperatures but also shows basic
similarities.  At the higher temperature, the wait time causes a
maximum of the relaxation rate at an observation time closely equal to
$t_{w}$.  At the lower temperatures, the corresponding maximum in the
rate curves is somewhat broader, of lower amplitude and somewhat
delayed compared to the actual $t_{w}$.  It should also be mentioned
here that the spin glass can be under-cooled or over-heated
arbitrarily large amounts, provided the cooling/heating rate is not
significantly altered; coming back to the measurement temperature
yields an almost identical response to what is observed when the
sample is directly brought to $T_{m}$.  The governing parameter for
the $t_{w}$=0 and the continued wait time dependence of the response
function, at the measurement temperature $T_{m}$, is the
characteristic time $t_{c}$ of the cooling rate.  The very fact that the
evolution or
ageing of the response function at vastly different temperatures shows
similar behaviour implies that the spin configuration that develops
during cooling differs at different temperatures; the system is
chaotic.  To carry this point further, we show results from an
experiment where the sample has been cooled to a temperature
$T_{m}\pm\Delta T$, kept there at constant temperature a wait time
3000 s and then cooled from $T_{m}$+$\Delta T$ (or heated from
$T_{m}$-$\Delta T$) to $T_{m}$, where the magnetic field is applied
after a short wait time that allows thermal stability to be attained.
The results of the experiment are shown in Fig.  3, where the
relaxation rate is plotted vs.  log $t$ for different magnitudes of
positive ($T_{m}$+$\Delta T$ ), Fig.  3 a, and negative
($T_{m}$-$\Delta T$), Fig.  3 b, temperature shifts $\Delta T$ .

Three different regions for the response can be distinguished:
Large $\Delta T$ : The response is indistinguishable from what is
measured if the sample is cooled directly to $T_{m}$ from a high
temperature above $T_{g}$(0), i.e.  the time that the sample has been
kept at constant temperature $T_{m}\pm\Delta T$ is irrelevant for the
response at $T_{m}$.  The spin configuration developed at
$T_{m}\pm\Delta T$ does not map onto the configuration that develops
at $T_{m}$.

Small $\Delta T$: The response appears unaffected by the temperature
shift, only the position of the maximum is pushed to shorter (negative
$\Delta T$) or longer (positive $\Delta T$) time than the actual wait
time.  This behaviour can be accounted for by the temperature
dependence of the dynamics and that there is an overlap between the
spin configurations at the two temperatures on the time scales of our
experiment.  Similar results have been interpreted using length scale
arguments \cite {nine} that also appear in droplet models.  It should
be noted that the behaviour for positive $\Delta T$ does not give the
perfect overlap with the corresponding response attained at $T_{m}$
that a negative $\Delta T$ gives.

\begin{figure}

\centerline{\hbox{\epsfig{figure=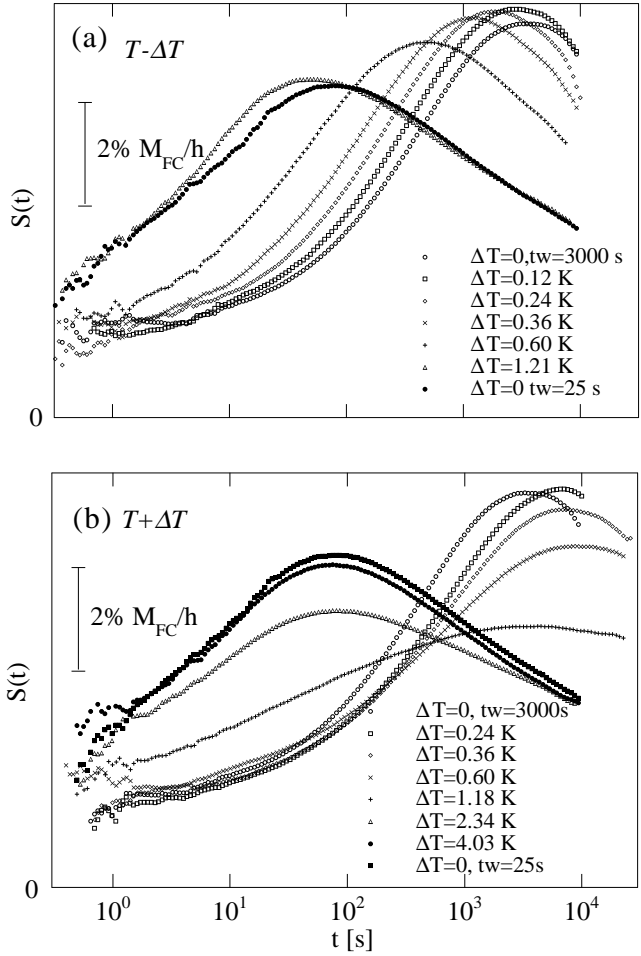,width=8.5cm}}}
\vskip 0.5cm
\caption{
\hbox  {The relaxation rate of the zero field cooled} magnetisation
measured after keeping the sample at constant temperature at
$T_{m}\pm\Delta T$
a wait time $t_{w}$=3000 s, and then shift the temperature to $T_{m}$= 58 K
and measure the relaxation of the magnetisation in an applied field,
$H$ = 1 Oe. a) shifts $T_{m}$-$\Delta T$ to $T_{m}$, b) shifts
$T_{m}$+$\Delta T$ to $T_{m}$.}
\label{fig3}
\end{figure}

Intermediate $\Delta T$: On positive temperature shifts, there is a
continuous shift of the position of the maximum which ends by closely
merging into the $t_{w}$=0 curve.

On negative temperature shifts the
behaviour is less pregnant and there is a dramatic broadening and
levelling off of the observable maximum that continuously changes to
end by perfectly merging into the $t_{w}$=0 curve for $\Delta T$ $>$
4-5 K.

These results show that in our time window, the overlap between the
attained configuration at the two temperatures is good for $\Delta T$
$<$ $\Delta T_{0}$ but gradually decreases with increasing magnitude
of $\Delta T$.  At large $\Delta T$'s there is no overlap, a chaotic
situation is observed.  It is however in this context also important
to recall results from temperature cycling experiments, i.e.
experiments where the spin glass is aged at $T_{m}$, then subjected to
a temperature cycling of magnitude $\Delta T$, and when $T_{m}$ is
recovered the magnetic field is applied and the relaxation of the

\begin{figure}
\centerline{\hbox{\epsfig{figure=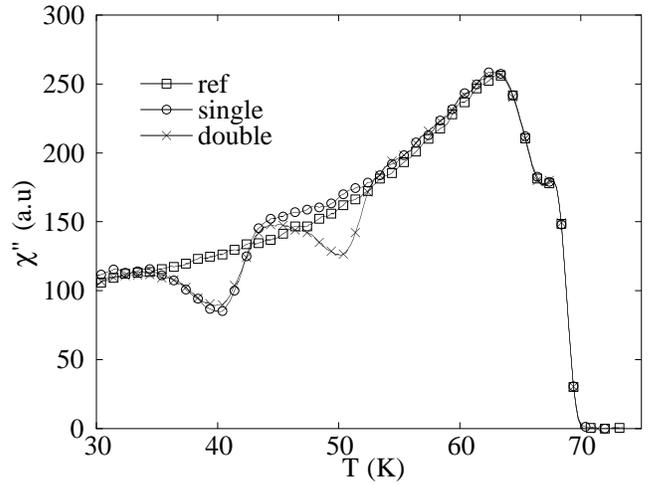,width=8.5cm}}}
\vskip 0.5cm
\caption{
\hbox  { $\chi''(T,\omega,t)$ vs. $T$. Singel memory experiment:}
while cooling, the sample was kept at $T_{1}$=40 K for 6 hours.
Double memory experiment: The sample was kept at $T_{1}$=50 K (6 hours) and
$T_{2}$=40 K (6 hours). Displayed in the figure are the resulting
curves achieved when the system is continuously reheated after the different
cooling procedures. The reference curve is measured on heating after
the system had been continuously cooled at a constant cooling rate.}
\label{fig4}
\end{figure}

\noindent magnetisation is recorded.  The temperature cycling experiments show
that on positive temperature cycles, the spin glass appears to have a
completely random initial state for large values of $\Delta T$,
whereas using negative $\Delta T$'s larger than $\Delta T_{0}$
requires a substantial wait time at the
lower temperature to achieve a
measurably reinitialised system \cite {ten}.  However, if even larger
negative values of $\Delta T$ are used, the effect of the cycling
again becomes small and the spin configuration attained at $T_{m}$
appears frozen.

Summarising these results on the properties of the response function,
we find: the response function and its wait time dependence is unique
for each temperature and cooling/heating rate.  When the wait time
becomes much longer than the time scale, $t_{c}$, of the
cooling/heating rate, the response function $p$($t$,$T$,$t_{w}$) is
determined only by $T$ and $t_{w}$.  If the sample is kept
intermittently at a constant temperature $T_{i}$ a time $t_{i}$, and
then further cooled and aged at a substantially lower temperature, the
response function when the temperature $T_{i}$ is recovered is closely
equal to only $p$($t$,$T_{i}$,$t_{i}$).  On the contrary, if the
temperature is increased substantially above $T_{i}$, the response
when $T_{i}$ is recovered is the same as cooling the sample directly
to $T_{i}$.  The system can carry the information from numerous
intermittent stops provided they are sufficiently separated in
temperature and that the temperatures are only recovered after heating
from lower temperatures.  This memory behaviour is also anticipated to
be observable in the temperature dependence of the ac-susceptibility
\cite {eleven}.  This is illustrated in Fig.  4 by a plot of the
temperature dependence of the out of phase component of the low
frequency, $f$=0.51 Hz, ac-susceptibility of our spin glass sample.
Three curves are displayed, which all are measured on heating at one
and the same constant heating rate after using three different cooling
procedures.  The curve marked ref. in the figure shows the result
when the sample has been cooled continuously to a low temperature.
The curve marked single shows the result when the sample was
intermittently halted for 6 hours at $T_{1}$=40 K during cooling.  The
curve marked double shows the result when the sample was
intermittently halted for 6 hours first at $T_{1}$=50 K and then at
$T_{1}$=40 K during cooling.  The result is in line with the
discussion above.  It is possible to retrieve information from not
only one stop but also two stops and more if they are well seperated
in temperature.  The dip(s) on re-heating the sample is a consequence
of the above discussed memory of the halt(s) at $T_{i}$ during
cooling.  The memory is erased when the temperature reaches well above
$T_{i}$.  Further experimental results illustrating this behaviour
have recently been published \cite {twelve}.

\section{Relaxation in constant field}

The field cooled magnetisation of spin glasses is a quantity that
falls out of equilibrium at the irreversibility temperature, i.e.  the
temperature where the ZFC and FC magnetisation curves merge.  At lower
temperature the FC magnetisation becomes cooling rate dependent and
relaxes if the sample is kept at constant temperature.  Some time
dependent features of the FC-magnetisation have been reported for 3d
spin glasses \cite {thirteen} and also for 2d spin glasses \cite
{fourteen}.  To connect to the experiments described above we have
performed measurements of the relaxation of the FC magnetisation after
cooling from a high temperature at one and the same cooling rate and
recording the relaxation at constant temperature.  Some curves
recorded at different temperatures are shown in Fig.  5.  It is worth
to note that in our experimental time window, there is an upward
relaxation at high temperatures near $T_{g}$, an initial downward
relaxation followed by a turning upwards on longer time scales at a
somewhat lower temperature and only a downward relaxation at lower
temperatures.  We choose to further investigate how the system
responds to a temperature cycling at the same measurement temperature
as the response function was studied.  In Fig.  6 a the results after
keeping the sample at 58 K and record the relaxation of the
magnetisation during 10$^{4}$ s, then subject the sample to a
temperature cycle of magnitude $\Delta T$, and when $T_{m}$ is
recovered record the continued relaxation of the magnetisation.  The
original magnitude and relaxation of the magnetisation is regained if
a large $\Delta T$ is used.  For $\Delta T$ $<$ $\Delta T_{0}$ the
relaxation continues apparently undisturbed when $T_{m}$ is recovered
and for intermediate values of $\Delta T$ a continuously increasing
regained magnitude of M is observed.  (The $\Delta T_{0}$ observed in
these experiments has about the

\begin{figure}
\centerline{\hbox{\epsfig{figure=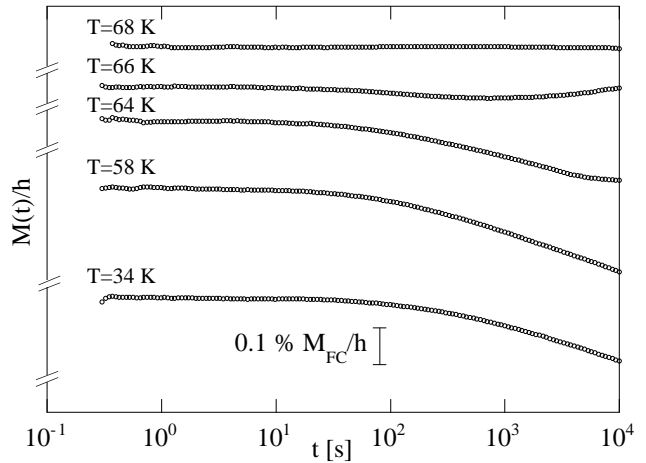,width=8.5cm}}}
\vskip 0.5cm
\caption{
\hbox  {  $M_{FC}$$(T_{m}$,$t$) vs. $\log$ $t$ measured at some 
different}
temperatures as
indicated in the figure.}
\label{fig5}
\end{figure}

\noindent same value as the $\Delta T_{0}$
obtained in the cycling experiments for the response function
recapitulated above.)  Fig.  6 b shows corresponding results using
negative cycling.  The relaxation of the magnetisation is first
recorded during 3000 s at $T_{m}$, then the sample is kept at the
lower temperature for 10$^{4}$ s and the relaxation is recorded.
Thereafter the sample is heated to $T_{m}$ and the continued
relaxation is recorded.  In the figure the relaxation is plotted at
$T_{m}$ also for $\Delta T$=0 to have a reference curve for the whole
measured time interval.  Some features are noticeable.  For small
values of $\Delta T$, the relaxation during the temperature cycle
results in a decreased magnetisation compared to only remaining at
$T_{m}$.  For larger values $\Delta T$ there is a slightly increased
magnetisation.  It should also be stressed that if the sample is
subjected to only a temperature cycling $\Delta T$ to a lower
temperature and heated back at the rate $t_{c}$, the magnetisation
retains the magnetisation attained after 3000 s and continues to relax
apparently unaffected by the temperature cycling.  The overall picture
is that the FC magnetisation is to some extent affected by the
cycling, but maybe most significantly, in spite of the rather long
wait time at the lower temperature, the magnetisation on recovering
$T_{m}$ is rather closely equal to the magnetisation level reached
when only keeping the sample at constant temperature.  Also seen from
the figure is that the relaxation during the period at the lower
temperature is in fact as large as if the sample had been immediately
cooled to that temperature without stopping at $T_{m}$, however with a
significantly decreased initial magnitude.  In summary it is found
that the field cooled magnetisation is essentially governed by:

(i) a closely reversible magnetisation,
$M_{FCrev}$($T$,$t_{c}$) is approached if the same cooling and heating
rate is used.

(ii) if the system is halted at a constant temperature, $T_{i}$, a
time $t_{i}$, the FC-magnetisation decays an amount, $\Delta
M$($T_{i}$, $t_{i}$) and if the cooling is continued at the original
cooling rate, the magnetisation remains at the lower

\begin{figure}
\centerline{\hbox{\epsfig{figure=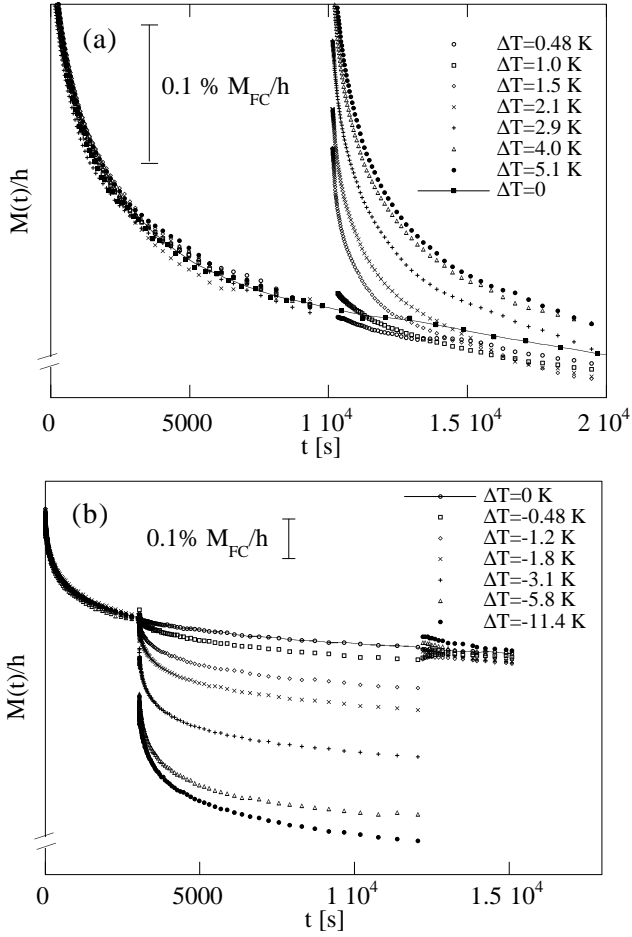,width=8.5cm}}}
\vskip 0.5cm
\caption{
\hbox  {  $M_{FC}$ vs. time at $T_{m}$ = 58 K and $H$ = 10 Oe.}
a) After 10$^{4}$ s the temperature is cycled $\Delta T$ and on recovering
$T_{m}$ the recording is continued. b) The relaxation at $T_{m}$
is recorded 3000 s, then the sample is cooled to $T_{m}$-$\Delta T$,
allowed to
relax during 10$^{4}$ s, as shown in the figure, and then heated
back to $T_{m}$ where the continued relaxation is recorded an additional
3000 s.}
\label{fig6}
\end{figure}

\noindent magnetisation
value:

\begin{equation}
M_{FC}(T>T_{i},t_{c})=M_{FCrev}(T,t_{c})
\end{equation}
\begin{equation}
M_{FC}(T<T_{i},t_{c})=M_{FCrev}(T,t_{c})-\Delta M(T_{i},t_{i})
\end{equation}

(iii) On re-heating the FC sample, the magnetisation rather adequately
follows the same track as during cooling, including an increase at the
temperature, $T_{i}$, where the sample was kept at constant
temperature (see Eq.  2).

These observations are summarised in the
schematic drawing of Fig.  7 where a cooling curve including
intermittent halts at two different temperatures and the subsequent
continuous heating curve are illustrated.  It is worth to notice that
the halt temperatures should be significantly separated for the
behaviour to be resolved.

Many experiments on relaxation in spin glasses are made using the
relaxation of the thermo remanent magnetisation
(TRM) as a probe, i.e. the relaxation observed

\begin{figure}
\centerline{\hbox{\epsfig{figure=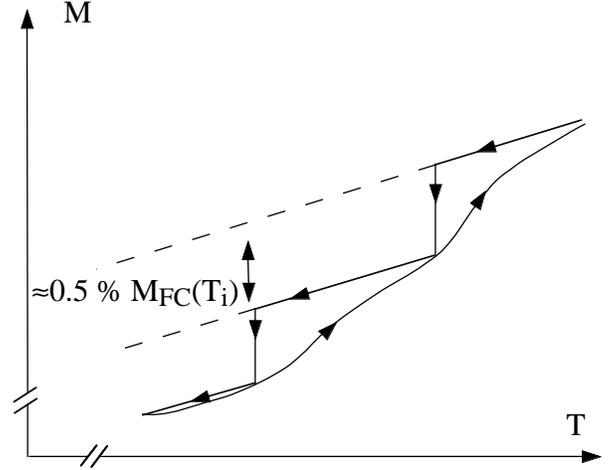,width=8.5cm}}}
\vskip 0.5cm
\caption{
\hbox  {  A schematic drawing of $M_{FC}$ vs. $T$ where two}
intermittent halts are made at two well separated temperatures during
cooling. The curve on continuously re-heating the sample is also
indicated. }
\label{fig7}
\end{figure}

\noindent after cooling the
sample in a constant field to a measurement temperature, cutting the
applied field and then recording the relaxation of the magnetisation.
The magnitude of the TRM
measured at a temperature well below an intermittent halt will have a
different magnitude than if immediately cooled to the measurement
temperature.  However, the relaxation rate will remain identical in
the two procedures.  For a $t_{w}$=0 example we can write:

\begin{equation}
M_{TRM}(t,t_{c})=M_{FCrev}(T,t_{c})+M_{ZFC}(t,t_{c})
\end{equation}
\begin{equation}
M_{TRM}(t,t_{c},t_{i})=(M_{FCrev}(t,t_{c})-\Delta M)+M_{ZFC}(t,t_{c})
\end{equation}

for the two cases.  That the termoremanent magnetisation carries
information on $M_{FC}$ distinguishes it from the zero field cooled
relaxation that always only sees the response function.

\section{conclusions}

From measurements of the response function and the FC magnetisation we derive
the following characteristics of the spin glass magnetisation and dynamics on
our experimental time scales.
An energetically favourable non-equilibrium spin configuration is imprinted on
a spin glass when cooling to a low temperature. This configuration is
essentially preserved when heating the sample at an identical heating rate,
the FC magnetisation is governed by the characteristic time $t_{c}$ and the
response function measured at constant temperature is independent of the
thermal history. If the sample is kept at a constant temperature, the spin
configuration rearranges on large length scales, which implies a decay of
the FC magnetisation and an altered response function. This configuration
is imprinted on the system and is 'frozen' in on lowering the temperature.
When the same temperature is recovered on re-heating, the sample appears as
if the thermal history at low temperature had not occurred.
These results verify that the spin glass phase is chaotic, that there is an
overlap between the equilibrium spin configurations developed at two slightly
different temperatures and also show that a spin configuration attained
at a high temperature is 'imprinted' on the system on lowering the
temperature.

\section{acknowledgments}

Financial support from the Swedish Natural Science Research Council
(NFR) is acknowledged.

\end{multicols}

\end{document}